# IMPROVING PERFORMANCE OF IEEE 802.11 BY A DYNAMIC CONTROL BACKOFF ALGORITHM UNDER UNSATURATED TRAFFIC LOADS

Hatm Alkadeki, Xingang Wang and Michael Odetayo

Department of Computing, Coventry University, Coventry, UK

## ABSTRACT

*The IEEE 802.11 backoff algorithm is very important for controlling system throughput over contention-based wireless networks. For this reason, there are many studies on wireless network performance focus on developing backoff algorithms. However, most existing models are based on saturated traffic loads, which are not a real representation of actual network conditions. In this paper, a dynamic control backoff time algorithm is proposed to enhance both delay and throughput performance of the IEEE 802.11 distributed coordination function. This algorithm considers the distinction between high and low traffic loads in order to deal with unsaturated traffic load conditions. In particular, the equilibrium point analysis model is used to represent the algorithm under various traffic load conditions. Results of extensive simulation experiments illustrate that the proposed algorithm yields better performance throughput and a better average transmission packet delay than related algorithms.*

## 1. INTRODUCTION

Use of wireless local area networks (WLANs) is ubiquitous. IEEE 802.11 is the most important standard in WLANs; however, achieving acceptable quality of service (QoS) over the standard is still a challenging task. The IEEE 802.11 standard provides a basic medium access control (MAC) mechanism called the distributed coordination function (DCF) [1]. The DCF is based on the carrier sense multiple access with collision avoidance (CSMA/CA) mechanism with binary exponential backoff (BEB) algorithm to reduce the probability of collisions. The BEB algorithm is implemented by doubling the backoff time after every unsuccessful transmission. The backoff time is called the contention window (*CW*), which is bounded by $CW_{max}$. However, the *CW* is reset to zero after every successful transmission with the backoff counter in the interval (0, $CW_i$ −1) [2]. Furthermore, the collision probability eventually leads to an unsuccessful transmission, which decreases throughput. Therefore, improving the backoff algorithm will help to enhance throughput performance and reduce the transmission delay. In this paper, we propose a new backoff algorithm that applies the distinction between low and high traffic loads over non-saturated traffic load conditions. We also evaluate the performance of the proposed algorithm in terms of throughput and delay.

The remainder of this paper is organized as follows. Section 2 discusses and evaluates related work, while Section 3 presents the proposed algorithm. Section 4 evaluates the performance of the proposed algorithm and compares the results with those of existing works. Section 5 presents our conclusion.

  45

## 2. RELATED WORK

As mentioned above, the backoff algorithm for IEEE 802.11 is very important for controlling channel access to maximize throughput and fairness [3]. There are several methods for extending or proposing backoff algorithms. Most of these are based on modifying the backoff parameters such as *CW* size and backoff stage (*m*), which is why much research has focused on modifying the *CW* size during execution of the backoff algorithm to improve the performance of the IEEE 802.11 DCF. Therefore, an appropriate *CW* size leads to an improvement in the system throughput by reducing the probability of collisions. However, some of the methods do not account for dynamic traffic loads. For example, the authors in [4], proposed a new backoff algorithm, called the multiplicative increase and linear decrease (MILD) algorithm. Their work focused on modifying the *CW* size to *CW*×1.5 rather than doubling it after every unsuccessful transmission. Moreover, *CW* size is decremented by 1 after every successful transmission rather than reset to 0. However, decreasing the *CW* size gradually helps avoid any degradation in performance. Therefore, the MILD algorithm is better than the BEB algorithm over large networks. The authors in [5], extended the MILD algorithm by creating a new algorithm called the linear increase linear decrease (LILD) algorithm. However, the authors applied *CW*+*CW*$_{min}$ as the size of increasing *CW* rather than multiplying by 1.5 to avoid the problem of slow linear change over unsuccessful transmission. Therefore, the LILD algorithm provides good quality performance over large networks. In another study [6], the authors proposed a new backoff algorithm, called the exponential increase exponential decrease (EIED) algorithm. This algorithm is based on increasing and decreasing the *CW* size exponentially. In [7], the authors proposed a new algorithm called the double increment double decrement (DIDD) algorithm. This algorithm is based on doubling the *CW* size after every unsuccessful transmission, in the same way as the BEB algorithm, but using *CW*/2 as the size of decreasing *CW* after every successful transmission. The DIDD algorithm generates a better result than the other algorithms mentioned above. In addition, improving the BEB algorithm is still an active research topic. Therefore, [8] recently evaluated the performance of BEB as a poor algorithm due to a number of collisions and *CW* restoration after every successful transmission. This study is devoted to improve collision avoidance under saturated traffic loads.

However, the above algorithms do not consider dynamic traffic loads. There are other interesting directions that can be taken. For example, according to the research in [9], the authors focused on channel traffic loads, and proposed a new algorithm called the exponential linear backoff algorithm (ELBA). ELBA combines both exponential and linear algorithms depending on traffic loads and provides better system throughput than the BEB, EIED, and LILD algorithms. In [10], the authors used pause count backoff for monitoring channel traffic loads. This algorithm aims to set an appropriate *CW* size based on estimation results. The authors in [11], proposed an adaptive backoff algorithm based on the trade-off of efficiency and fairness for ad hoc networks. This work is based on a fair schedule to control the increase and decrease in *CW* size depending on the channel situation (idle or busy). In [12], the authors considered dynamic traffic loads by proposing an algorithm based on monitoring the channel before data transmission. In this algorithm, each station can record the number of busy slots by opening an observation window. Thus, the sender can calculate a dynamic priority and *CW* size according to the number of successful transmissions. In [13], the authors monitored the channel traffic loads by using a channel state (CS) vector, and proposed a new algorithm called the dynamic deterministic contention window control algorithm (DDCWC). This algorithm is based on monitoring the channel traffic load conditions by checking the CS. However, selecting the optimum *CW* size based on different traffic load conditions using the CS vector is difficult.

Overall, the majority of research work has paid great attention to improving the performance of a saturated system without accounting for non-saturated traffic load conditions. Therefore, creating a new backoff algorithm under non-saturated traffic load conditions is the objective of this paper.

## 3. BACKOFF STRATEGY

In this section, the proposed algorithm is discussed in detail. The discussion starts by describing the principle behind the proposed algorithm in terms of mechanism and traffic load conditions.

### 3.1 Principle of the Proposed Algorithm

As mentioned in Section 2, most existing algorithms do not consider traffic loads under non-saturated conditions, and thus do not take into account practical network operation. In this section, a new backoff algorithm is proposed, called the dynamic control backoff time algorithm (DCBTA). The DCBTA is implemented under non-saturated traffic loads using the equilibrium point analysis (EPA) model [14]. The EPA model provides a very convenient way of evaluating system performance under non-saturated traffic loads, thereby enabling the presentation of the DCBTA under more flexible traffic sources. Furthermore, it is possible to investigate network traffic load conditions under a different number of stations.

In the DCBTA, channel conditions are checked by a $CW$ threshold ($CW_{Threshold}$). The $CW_{Threshold}$ value serves as a reference point for the collision rate. Therefore, $CW_{Threshold}$ plays a major role in the proposed algorithm as illustrated in Figure 1. The $CW_{Threshold}$ size is dependent on the maximum contention window size ($CW_{max}$), where the value of $CW_{Threshold}$ is equal to half that of $CW_{max}$. For example, the value of $CW_{max}$ in [14] was selected to be 1024. In this case, the value of $CW_{Threshold}$ is set to 512. Each state of node "$i$" is initially set to the minimum $CW$ size ($CW_{min}$), which can be increased up to $CW_{max}$.

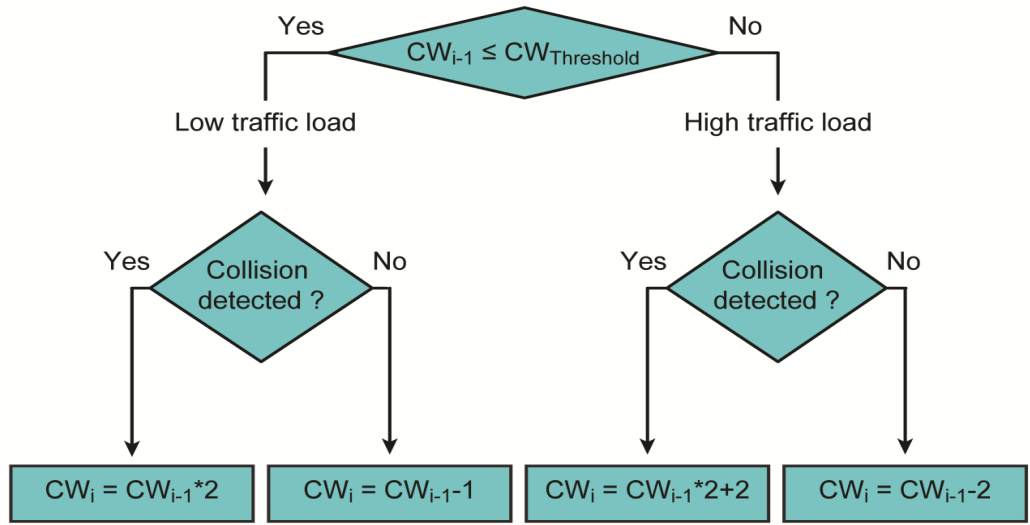

Figure 1. Underlying mechanism of the proposed algorithm (DCBTA)

Figure 1 shows that the proposed algorithm enables the detection of heavy or light traffic load using the $CW_{Threshold}$ value. After every unsuccessful transmission, if the $CW$ size is smaller than the $CW_{Threshold}$ value, that is, a light traffic load, the $CW$ size is doubled as ($2 \times CW$) similar to the BEB algorithm. Conversely, if the $CW$ size is greater than $CW_{Threshold}$, that is, a heavy traffic load, the $CW$ size is doubled and incremented by two as ($2 \times CW + 2$). Adding two to double the $CW$ size leads to a decrease in the collision probability, thus increasing system throughput. A summary of this discussion is given below:

International Journal of Wireless & Mobile Networks (IJWMN) Vol. 7, No. 6, December 2015

- **Light traffic load:**
  If ($CW_i \leq CW_{Threshold}$)
  Successful transmission: $CW_i = CW_{i-1} - 1$;
  Else ($CW_i = CW_{i-1} \times 2$).

- **Heavy traffic load:**
  If ($CW_i > CW_{Threshold}$)
  Successful transmission: $CW_i = CW_{i-1} - 2$;
  Else ($CW_i = CW_{i-1} \times 2 + 2$).

### 3.2 DCBTA Algorithm under EPA Model

In order to run the proposed algorithm under non-saturated traffic load conditions, the EPA model is used. The EPA model provides a very convenient way to evaluate the system performance under non-saturated traffic load conditions. In the EPA model, the traffic load generated by each station follows the Poisson distribution with rate time/packets. Thus, the packet transmission probability ($R$) plays a pivotal role in the EPA model mechanism. However, the proposed algorithm adaptively changes the $CW$ size with respect to the collision rate or the transmitting nodes. Therefore, the proposed algorithm under the EPA model affects the transmission probability of node "$R_i$" at any state of node "$i$" as follows:

$$R_i = \frac{1}{CW_i}$$

In networks with a large number of nodes or a high collision rate, the proposed algorithm results in a very low probability of transmission. In this case, the $CW$ size increases to more than the threshold, resulting in a high traffic load. The throughput formula is the same, where $R_i$ is calculated as follows:

$$R_i = \frac{1}{2^i CW_{min} + 2}$$

Otherwise, the value of $CW_i$ decreases to less than or equal to the threshold value, resulting in a low traffic load. Then $R_i$ is calculated in the same way as the BEB algorithm under the EPA model. In the case of successful transmission, the $CW_i$ size decreases gradually to avoid performance degradation. However, if the $CW_i$ size is less than or equal to $CW_{Threshold}$, the $CW$ size for the next stage $CW_i+1$ is decremented by one:

$$CW_i = CW_{i-1} - 1$$

If $CW_i$ is greater than $CW_{Threshold}$, $CW_i+1$ is decremented by two:

$$CW_i = CW_{i-1} - 2$$

### 4. PERFORMANCE EVALUATION

In this section, the proposed backoff algorithm is compared with related algorithms in terms of throughput and average packet transmission delay. The comparative evaluation of backoff algorithms is carried out using MATLAB simulation experiments.

48

### 4.1 Simulation Settings

The proposed and related algorithms are implemented based on the EPA model assumption in [14]. Therefore, there are no hidden terminals and system performance can be investigated under more flexible traffic sources with fixed packet length. The different system parameters used in the simulation experiments are summarised in Table 1.

Table 1. System parameter settings

| Parameter | Value |
| --- | --- |
| Packet Payload | 8184 bits |
| Data Packet | 8200 μs |
| Channel Bit Rate | 1 Mbit/s |
| Physical Slot Time | 50 μs |
| DIFS | 128 μs |
| SIFS | 28 μs |
| ACK_Timeout | 300 μs |
| RTS | 350 μs |
| CTS | 350 μs |
| $CW_{min}$ | 8 |
| $CW_{max}$ | 1024 |
| Maximum Backoff Stage, $m$ | 6 |
| Network Nodes ($n$) | 50-100 nodes |
| Analytical Tool | EPA model |

### 4.2 Comparison of Throughput

System performance of the proposed algorithm (DCBTA) is compared with that of the BEB algorithm under non-saturated traffic load conditions in the work of [14]. In addition, the performance of DCBTA is compared with other related algorithms, such as ELBA in the work of [9]. ELBA combines both exponential and linear algorithms, which is why it was selected for comparison with the proposed algorithm. The number of nodes is set to 50; the maximum number of backoff stages equals six. Figure 2 illustrates the throughput performance for DCBTA compared with the BEB algorithm and ELBA under various traffic load conditions. The results show that the throughput performance of DCBTA is better than that of the BEB algorithm and ELBA under various traffic loads.

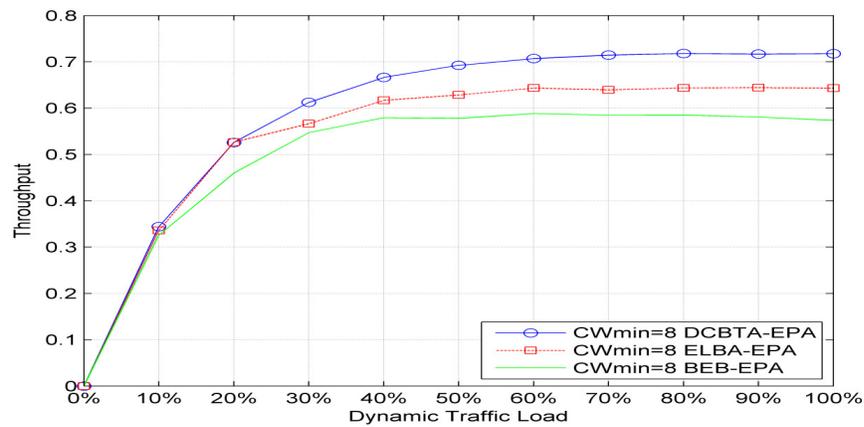

Figure 2. Non-saturated throughput comparison of the proposed algorithm and various related backoff algorithms, where ($CW_{min}$ = 8, $m$ = 6)

To investigate the impact of using different $CW_{min}$ size, Figure 3 plots the throughput performance for DCBTA, BEB, and ELBA with a varying size of 8, 16, and 32 $CW_{min}$. The throughput increases when $CW_{min}$ increases, since increasing $CW_{min}$ contributes to collision avoidance. Moreover, system throughput depends on the incoming data [15]. Therefore, the throughput result is equal to the increase in the incoming traffic data rates if the traffic load is low. Otherwise, throughput becomes saturated if the amount of data is sufficiently high. Hence, the system performance strongly depends on system parameters, such as $CW_{min}$ and $m$.

Figure 3 clearly shows that DCBTA provides better throughput results than BEB and ELBA with different $CW_{min}$ size under various offered loads. The DCBTA allows the stations to adjust $CW$ value appropriately according to the traffic load variation within the network. This means that the DCBTA mechanism can reduce the number of collisions, which will lead to increased system throughput. In addition, the performance results show that DCBTA has lower performance degradation than BEB and ELBA. The reason for this is that the $CW$ size decreases gradually after every successful transmission.

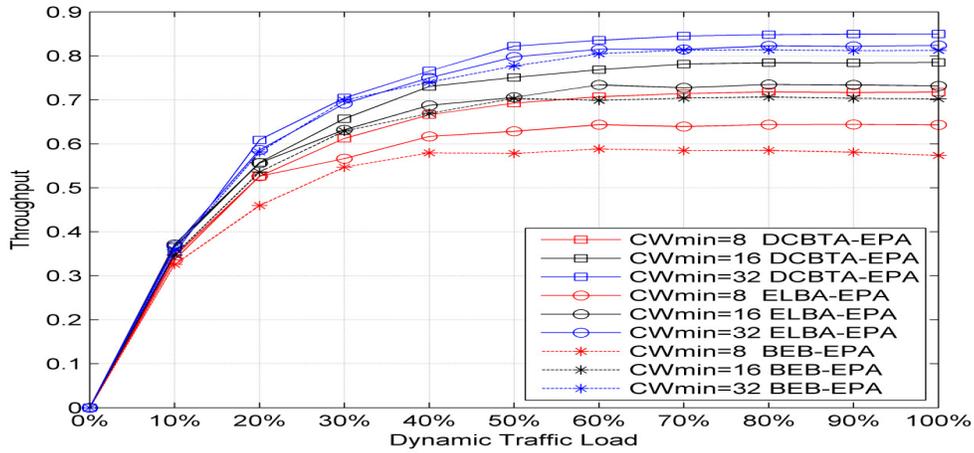

Figure 3. Non-saturated throughput comparison of the proposed algorithm and various related backoff algorithms with varying $CW_{min}$ (8, 16, 32) and $m = 6$

### 4.3 Comparison of Delay

In [14], the EPA model represented the MAC channel in idle, transmission, and collision states under varying traffic load conditions. The MAC channel was proposed as a multi-dimensional discrete-time Markov chain analysis model. Therefore, the delay can be represented as a sequence of discrete time delays as follows:

Average transmission delay = Total delay / Total number of transmissions,

where:

Total delay = Total transmission time + Total time delay in the collision + Backoff time + Empty slot.

Total transmission time = Transmission time of single packet × Total number of transmissions.
Transmission time of single packet = RTS + SIFS + CTS + SIFS + Data + SIFS + ACK + DIFS.
Total time delay in the collision = Delay time of single collision × Total number of collisions.
Delay of single collision = RTS + DIFS.

Average packet transmission delays for the BEB algorithm, ELBA, and DCBTA are calculated over 100 stationary nodes. For further investigation, the performance of algorithms is also examined under different $CW_{min}$ values of 32, 64, and 128. All the assumptions and system parameters related to this experiment are the same as in the previous section. Figure 4, Figure 5, and Figure 6 show the delay comparison of the BEB, ELBA, and DCBTA algorithms under the EPA unsaturated model. The increment in *CW* size in the BEB and ELBA algorithms results in greater delay compared to that of the DCBTA algorithm. This means that the DCBTA mechanism produces a small delay by reducing a collision rate. Actually, when there is a high offered traffic load, the *CW* size should be kept large to avoid frequent collision. Moreover, DCBTA reduces *CW* size more slowly after successful transmission in order to avoid the collision probability. For these reasons, it can clearly be seen that the proposed algorithm has a smaller average transmission delay than that of the BEB and ELBA algorithms, as shown in the figures below.

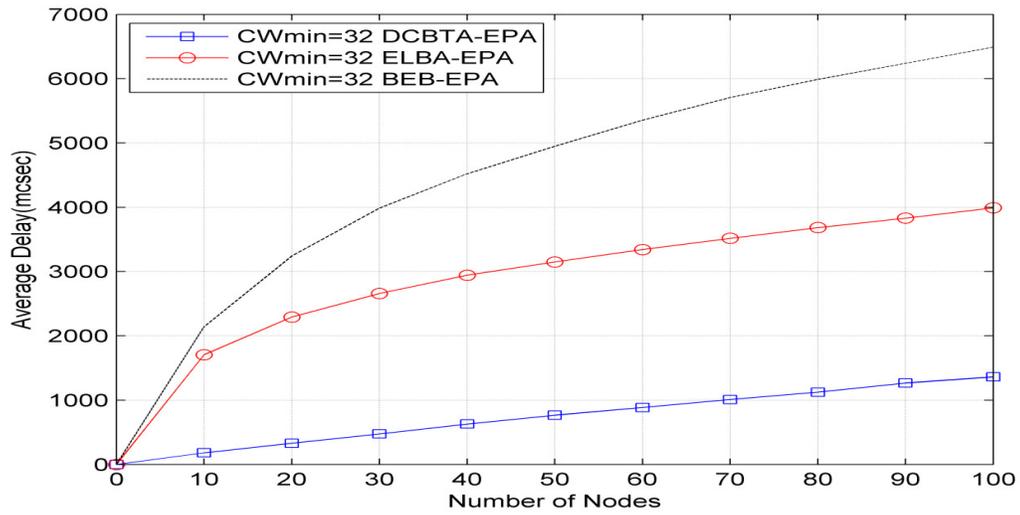

Figure 4. Average packet transmission delay with $CW_{min} = 32$, $m = 6$

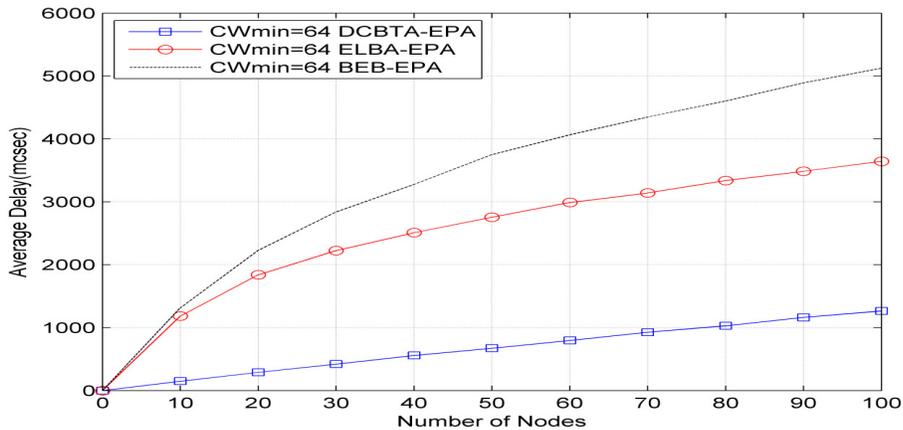

Figure 5. Average packet transmission delay with $CW_{min} = 64$, $m = 6$

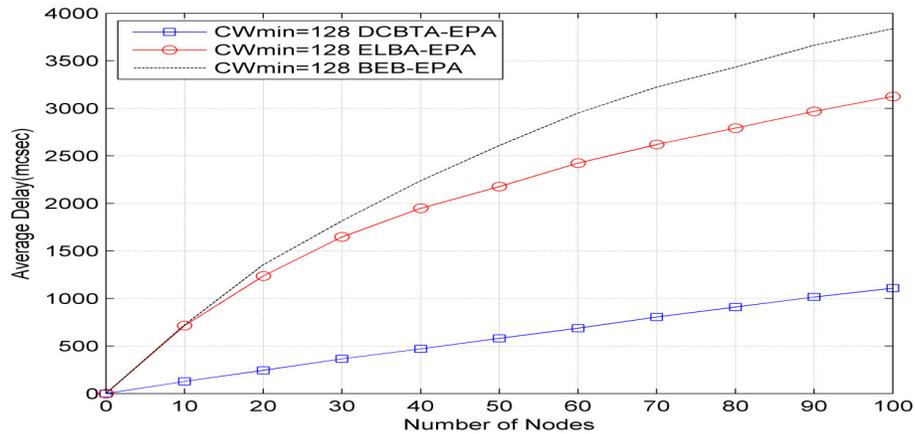

Figure 6. Average packet transmission delay with $CW_{min}$ = 128, $m$ = 6

## 5. CONCLUSION

In this paper, a new backoff algorithm under non-saturated traffic loads was proposed to represent actual network situations. A suitable model was selected to evaluate system performance under non-saturated traffic loads such as the EPA model.

The motivation for this research was to enhance the system performance of the IEEE 802.11 DCF under non-saturated traffic load conditions in terms of throughput and time delay. To realize this, a new backoff algorithm was proposed and then integrated with the EPA model.

The performance results show that the proposed algorithm (DCBTA) presents better system throughput than the BEB algorithm and ELBA. In addition, calculation of the average packet transmission delay for each algorithm shows that the DCBTA provides a better time delay than the BEB algorithm and ELBA. This is because the DCBTA decreases the time delay, which leads to an increase in system throughput. However, throughput and delay are both relevant for the performance metrics of QoS. Therefore, the proposed algorithm may help to enhance the effectiveness of the IEEE 802.11 DCF. A possible further extension of the DCBTA would be to consider the various data frame sizes.